\documentclass{PoS}
\usepackage{amsmath}
\usepackage{slashed}
\usepackage{subfigure}

\newcommand{\beq}{\begin{equation}}
\newcommand{\eeq}{\end{equation}}

\newcommand{\LSB}{\raisebox{-0.3ex}{\mbox{\LARGE$\left[\right.$}}}
\newcommand{\RSB}{\raisebox{-0.3ex}{\mbox{\LARGE$\left.\right]$}}}

\def\gsim{\mathrel{\raise2pt\hbox to 8pt{\raise -5pt\hbox{$\sim$}\hss{$>$}}}}
\def\rsim{\mathrel{\raise2pt\hbox to 8pt{\raise -5pt\hbox{$\sim$}\hss{$>$}}}}
\def\lsim{\mathrel{\raise2pt\hbox to 8pt{\raise -5pt\hbox{$\sim$}\hss{$<$}}}}

\title{B-physics computations from $N_f=2$ tmQCD}

\ShortTitle{B-physics computations from $N_f=2$ tmQCD}

\author{ N. Carrasco$^{(a)}$, M. Ciuchini$^{(b)}$, \speaker{P. Dimopoulos}$^{(c,d)}$, R. Frezzotti$^{(d)}$, V. Gim\'enez$^{(a)}$  
G. Herdoiza$^{(f)}$, V. Lubicz$^{(g,b)}$,  C. Michael$^{(h)}$, E. Picca$^{(g)}$, G.C. Rossi$^{(d)}$, 
F. Sanfilippo$^{(i)}$, A.~Shindler$^{(j)}$, L. Silvestrini$^{(k)}$, 
S. Simula$^{(b)}$, C. Tarantino$^{(g,b)}$ \\
 $^{(a)}$  Departament de F\'{\i}sica Te\`orica and IFIC, Univ. de Val\`encia-CSIC  
 Dr.~Moliner 50, E-46100 Val\`encia, Spain,  E-mail: \email{nuria.carrasco@uv.es, vicente.gimenez@uv.es}\\
 $^{(b)}$ INFN, Sezione di Roma Tre,c/o Dipartimento di Fisica, Universit\`a  Roma Tre, Via della Vasca Navale 84, I-00146 Rome, Italy, 
 E-mail: \email{  \{ciuchini, simula\}@roma3.infn.it} \\
  $^{(c)}$ Centro Fermi - Museo Storico della Fisica e Centro Studi e Ricerche Enrico Fermi, 
 Compendio del Viminale,  Piazza del Viminale 1, I-00184 Rome, Italy,\\
 $^{(d)}$ Dipartimento di Fisica, Universit\`a di Roma ``Tor Vergata'' and INFN Sezione ``Tor Vergata", \\ 
 Via della Ricerca Scientifica 1, I-00133 Rome, Italy, \\
 E-mail: \email{\{dimopoulos,frezzotti,rossig\}@roma2.infn.it} \\
 $^{(f)}$ PRISMA Cluster of Excellence, Institut f{\"u}r Kernphysik,Johannes Gutenberg-Universit{\"a}t, D-55099 Mainz, Germany,
  E-mail: \email{herdoiza@kph.uni-mainz.de} \\
  $^{(g)}$ Dipartimento di Fisica, Universit\`a  Roma Tre, Via della Vasca Navale 84, I-00146 Rome, Italy,
 E-mail: \email{e.picca88@gmail.com, \{lubicz, tarantino\}@fis.uniroma3.it,}\\
  $^{(h)}$ Theoretical Physics Division, Department of Mathematical Sciences, The University of Liverpool, Liverpool L69 3BX, UK,
   E-mail: \email{C.Michael@liverpool.ac.uk} \\
  $^{(i)}$ Laboratoire de Physique Th\'eorique (B\^{a}t. 210), Universit\'e Paris Sud, F-91405 Orasay-Cedex, France, 
  E-mail: \email{francesco.sanfilippo@th.u-psud.fr} \\
  $^{(j)}$ CERN, Physics Department, 1211 Geneva 23, Switzerland \\ 
  $^{(k)}$ INFN, Sezione di Roma, c/o Dipartimento di Fisica, Sapienza, Universit\`a di Roma, Piazzale A. Moro, I-00185 Rome, Italy, 
   E-mail: \email{Luca.Silvestrini@roma1.infn.it} 
 }

\abstract{We present an accurate  lattice QCD computation of the $b$-quark mass, the $B$ and $B_s$ decay constants, 
the $B$-mixing bag-parameters for the full four-fermion operator basis, as well as estimates for
$\xi$ and $f_{Bq}\sqrt{B_q}$ extrapolated to the continuum limit and the physical  pion mass. 
We have used $N_f = 2$ dynamical quark gauge configurations
at four values of the lattice spacing generated by ETMC. 
Extrapolation in the heavy quark mass from the charm to the bottom quark region has been carried
out using ratios of physical quantities computed at nearby quark masses, having an exactly known infinite mass limit.}

\FullConference{31st International Symposium on Lattice Field Theory - LATTICE 2013\\
		July 29 - August 3, 2013\\
		Mainz, Germany}

\begin{document}

\section{Introduction}
Many precision tests of the Standard Model (SM) and some proposals regarding stringent constraints on 
New Physics (NP) generalisations of the SM involve accurate studies of processes 
in the $b$-quark region. Experimental measurements on the leptonic decays 
$B \rightarrow \tau \nu_{\tau}$ and $B_{(s)}^0 \rightarrow \mu^{+} \mu^{-}$
need precise input information of the 
pseudoscalar decay constants $f_B$ and $f_{Bs}$ from lattice computations. 
Neutral $B$-meson oscillations are described, in the SM,  by a single four-fermion operator whose matrix element (bag parameter) can be 
determined in lattice simulations. $B$-meson mixing, besides its crucial role in the Unitarity Triangle (UT) analysis, can
provide important clues for detecting NP effects. The knowlegde of the values of the bag parameters 
of the complete four-fermion operator basis is required to obtain predictions about the NP scale for physics beyond the SM. 

In these proceedings we report on computations of a number of $B$-physics hadronic observables 
performed within the European Twisted Mass Collaboration (ETMC). 
We have computed in the continuum limit, by making use of data at four values of the lattice spacing, 
the value of the $b$-quark mass, the pseudoscalar meson decay constants, $f_B$ and $f_{Bs}$ as well as their ratio. 
Moreover we have  determined  the bag parameters corresponding to the full four-fermion operator basis,
as well as  other interesting phenomenological quantities like  $\xi$ and $f_{Bq}\sqrt{B_i^{(q)}}$ ($i=1, \ldots, 5$ and $q=d/s$).
For a more detailed discussion of methods and results of this study the reader is referred to a 
recent ETMC publication~\cite{Carrasco:2013zta} as well as to~\cite{BP-older1, BP-older2}. 
First ETMC results of the $b$-quark mass using 
$N_f=2+1+1$ gauge ensembles can be found in~\cite{LAT2013-2+1+1}.

\section{Lattice setup and computational details}
We have employed $N_f=2$ dynamical quark gauge configurations  
generated by ETMC~(\cite{etmc-nf2-1} -- \cite{etmc-nf2-3})  using Wilson twisted mass action tuned to maximal twist~\cite{F-R-1}
at four values of the inverse bare gauge coupling,
$\beta$. The lattice spacing values lie in the interval [0.05, 0.1] fm.
Light quark mass values of the degenerate $u/d$ quark produce light pseudoscalar mesons (``pions") in the range
$280 \leq m_{\rm PS} \leq 500$~MeV. 
We recall that the maximally twisted fermionic action offers the advantage  of automatic O$(a)$  improvement for
all the physical observables computed on the lattice~\cite{F-R-1}.
Strange and charm quarks  are treated in the quenched approximation. 
We have computed 2- and 3-point correlation functions 
using valence quark masses whose range is extended from the light sea quark mass up to 2.5-3 times the charm quark mass.
The values of the  light valence quark mass  are set equal
to the light sea ones.
Renormalised quark masses are determined from the bare ones
using the renormalisation constant (RC)
$Z_{\mu} = Z_P^{-1}$,  $\mu^{R}=\mu^{bare} / Z_P$.
The $Z_P$ values have been computed using RI/MOM techniques in~\cite{Constantinou:2010gr}.
Details of the ETMC determinations of light, strange and charm quark mass using 
$N_f=2$ gauge ensembles are given in~\cite{Blossier:2010cr}.

We have employed smeared interpolating operators in combination with APE smeared gauge links to compute 
2- and 3- point correlation  functions. With respect to the local fields the use of smeared ones reduce the overlap 
with the excited states and increased the projection onto the lowest energy eigenstate. 
We are able thus to extract heavy-light meson masses and matrix 
elements at relatively small Euclidean time separations. 
Moreover, as it is shown in~\cite{Carrasco:2013zta}, even better projection onto the ground state is achieved 
by employing a linear combination of smeared and local interpolating fields. The optimal linear combination  is 
determined by tuning the superposition parameter.

\section{Computation of the $b$-quark mass and decay constants $f_B$ and $f_{Bs}$}
\begin{figure}[!ht]
\begin{center}
\subfigure[]{\includegraphics[scale=0.62,angle=-0]{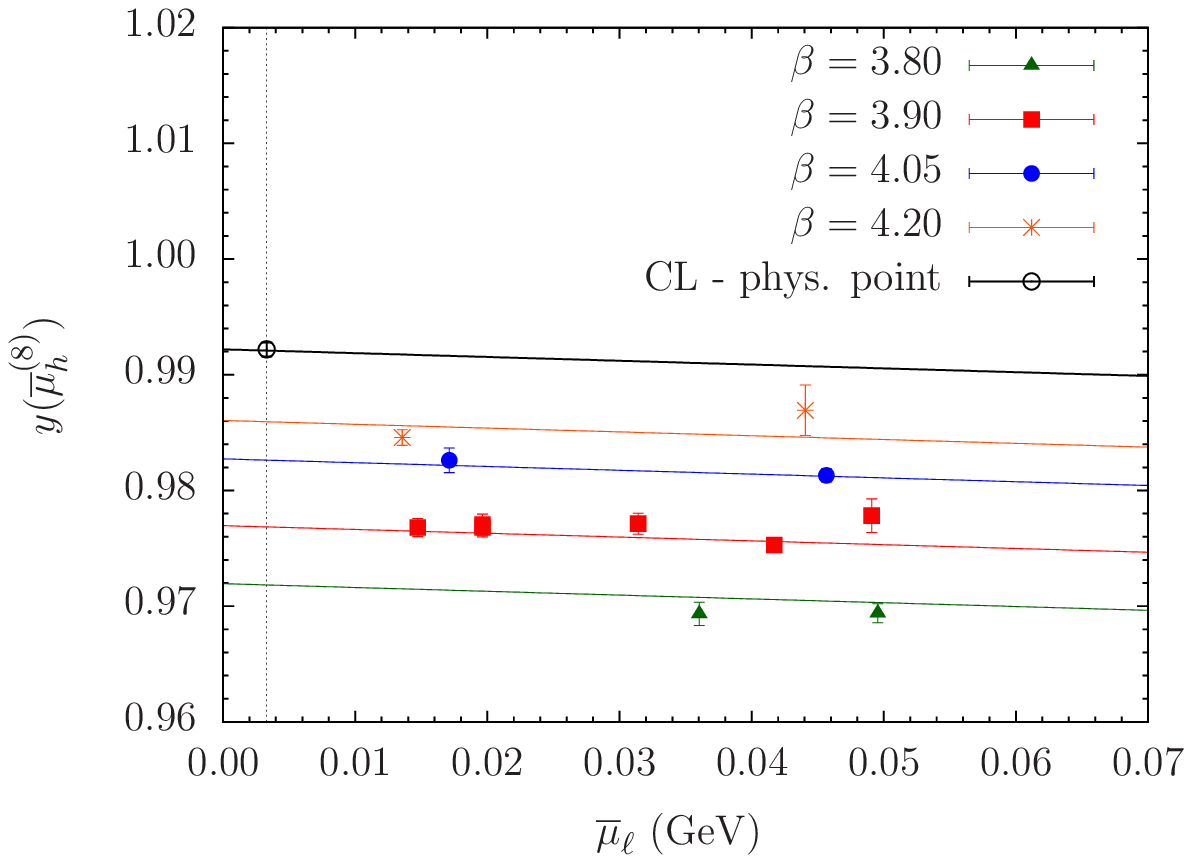}}
\subfigure[]{\includegraphics[scale=0.62,angle=-0]{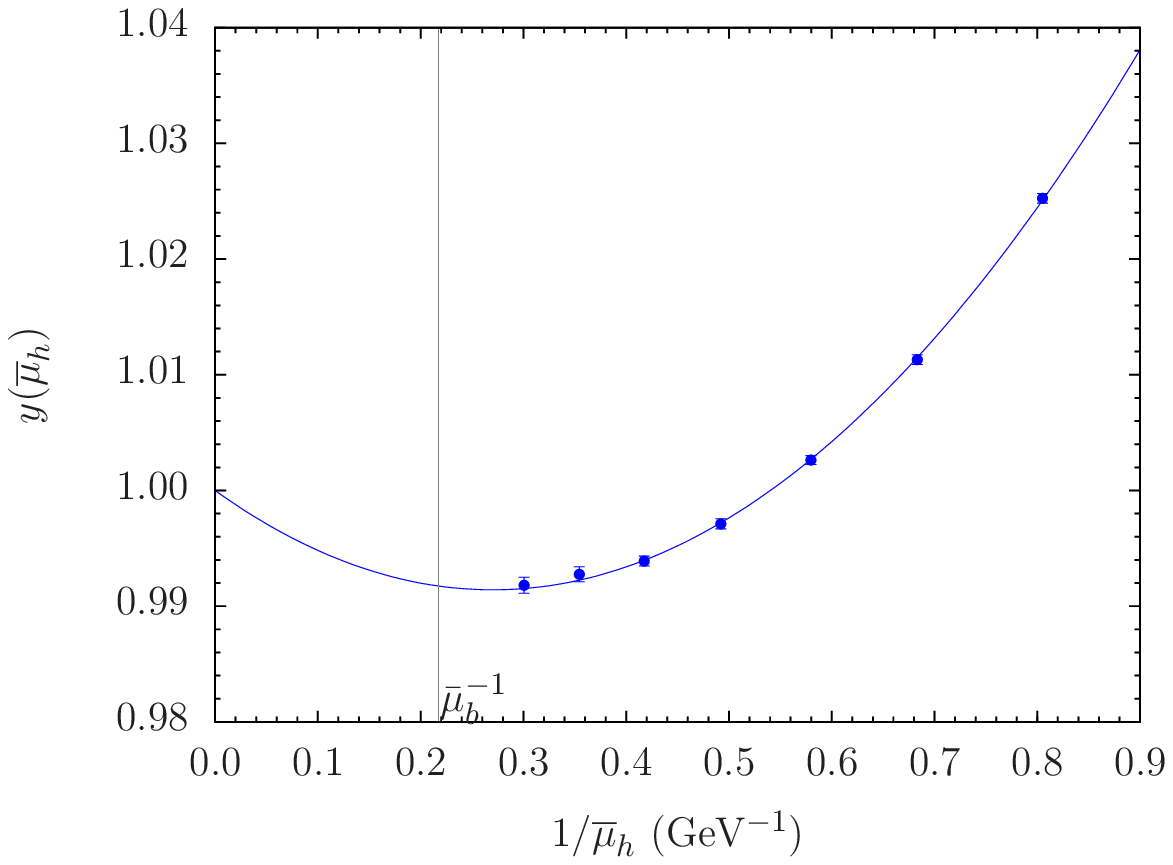}}
\vspace*{-0.3cm}
\caption{(a) Combined chiral
and continuum fit of the ratio of heavy-light pseudoscalar meson mass 
for our largest value of the heavy quark mass (empty black circle is our estimate at the physical $u/d$ quark mass 
in the continuum
limit). (b) $y(\overline\mu_h)$ against $1/\overline\mu_h$ using the fit ansatz~(3.3)
($\chi^2 / d.o.f.=0.3$).  }
\label{fig:y}
\end{center}
\end{figure}

For the determination of $B$-physics observables for which the asymptotic behaviour is known from HQET, it is in general possible 
to use the ratio method described in~(\cite{Carrasco:2013zta} -- \cite{BP-older2}). We briefly recall the computation for the $b$-quark mass.
HQET suggests that in the static limit the heavy-light pseudoscalar 
meson mass obeys 
\begin{equation}
\lim_{\mu_h^{\rm{pole}}\to \infty}  \left(M_{h\ell}/\mu_h^{\rm pole}\right) = \rm{const.} 
\label{eq:Mhl}
\end{equation}
We consider a set of heavy quark masses $(\overline\mu_h^{(1)}, \overline\mu_h^{(2)},  \cdots, \overline\mu_h^{(N)})$ 
having a fixed ratio, $\lambda$, between any two successive values: $\overline\mu_h^{(n)} = \lambda \overline\mu_h^{(n-1)}$. 
Throughout this work the ``overline" notation for the quark masses denotes that they are renormalised  
in the $\overline{\rm{MS}}$ scheme at 3 GeV.
We construct properly normalised ratios namely, $y(\overline\mu_h^{(n)},\lambda;\overline\mu_{\ell},a)$,  
of the quantity in the lhs of Eq.~(\ref{eq:Mhl}) at nearby values of the heavy quark mass. 
Taking the static and continuum limit one  gets the exact equation
\begin{equation}
\lim_{\overline\mu_h \to \infty} \lim_{a \to 0}\, y(\overline\mu_h^{(n)},\lambda;\overline\mu_{\ell},a) = 1\, . 
 \end{equation} 
At each value of $\overline\mu_h^{(n)}$ we perform a combined chiral and continuum fit on the ratios to extract the quantity 
$y(\overline\mu_h) \equiv y(\overline\mu_h, \lambda;\overline\mu_{u/d},a=0)$. Discretisation effects in ratios are well under control. 
In Fig.~\ref{fig:y}(a) we show the data and the fit for the ratio that corresponds to the largest simulated 
pair of the heavy quark mass values.
We describe the dependence of the ratio $y(\overline\mu_h)$ on $\overline\mu_h$ using the  fit ansatz
\begin{equation} \label{eq:y-ansatz}
y(\overline\mu_h) = 1 + \dfrac{\eta_1}{\overline\mu_h} + \dfrac{\eta_2}{\overline\mu_h^2},
\end{equation}
in which the constraint $\lim_{\overline{\mu}_h \rightarrow \infty} y(\overline\mu_h) = 1$ is implemented.
The fit parameters could be, in general,
functions of $\log(\overline\mu_h)$. However in the range of the
currently explored heavy quark mass values this logarithmic dependence can be safely neglected. 
Data and fit are shown in Fig.~\ref{fig:y}(b). The final step consists in determining the $b$-quark mass value  from the
{\it chain} equation
\begin{equation}\label{eq:y_chain}
   y(\overline\mu_h^{(2)})\, y(\overline\mu_h^{(3)})\,\ldots \, y(\overline\mu_h^{(K+1)})=\lambda^{-K} \,
\frac{M_{hu/d}(\overline\mu_h^{(K+1)})}{M_{hu/d}(\overline\mu_h^{(1)})} \cdot
\Big{[}\frac{\rho( \overline\mu_h^{(1)},\mu)}{\rho( \overline\mu_h^{(K+1)},\mu)}\Big{]}\,,
\end{equation}
where the function $\rho( \overline\mu^{(n)}_h,\mu)$, that is known up to N$^3$LO in perturbation theory, 
relates the $\overline{\rm{MS}}$ renormalised quark mass (at the scale of $\mu=3$ GeV)
to the pole mass: $\mu_h^{\rm{pole}} = \rho(\overline\mu_h, \mu) \, \overline\mu_h(\mu)$.
In Eq.~(\ref{eq:y_chain}) $\lambda$, $K$ and $\overline\mu_h^{(1)}$ have been chosen in such a way 
that $M_{hu/d}(\overline\mu_h^{(K+1)})$ coincides with
the experimental value of the $B$-meson mass, $M_{B}=5.279$ GeV. 
The quantity $M_{hu/d}(\overline\mu_h^{(1)})$ is the result of the combined chiral and continuum fit
of  pseudoscalar meson mass values evaluated 
at the reference heavy quark mass, $\mu_h^{(1)}$, that lies close to the charm quark mass.  
For $(\overline\mu_h^{(1)}, \lambda) = (1.05~\rm{GeV},~ 1.1784)$, Eq.~(\ref{eq:y_chain}) is satisfied for $K=K_b=9$. 
The $b$-quark mass in the $\overline {\rm{MS}}$ scheme at 3 GeV is
\begin{equation}\label{eq:mb}
 m_b(m_b, \overline{\rm{MS}}) = 4.29(9)(8)[12] \,\, {\rm GeV}, 
 \end{equation}
where the first two errors are statistical and systematic, respectively. 
Adding them in quadrature gives the total error in the square brackets. 
Having calculated the value of the $b$-quark mass, we follow analogous strategies to compute the heavy-strange pseudoscalar 
decay constant, $f_{Bs}$, and the value of the ratio of the 
heavy-strange and the heavy-light decay constants, $f_{Bs}/f_{B}$. We then get  our estimate for the heavy-light pseudoscalar 
decay constant from $f_B = f_{Bs}~/~(f_{Bs}/f_B)$. Statistical errors are estimated using the bootstrap method. 
Our results read  
 \begin{equation}\label{eq:fB}
 f_{Bs} = 228(5)(6)[8]\,\, {\rm MeV}, \,\,\,\, f_{B}  = 189(4)(7)[8]\,\, {\rm MeV},\,\,\,\,  f_{Bs}/f_B = 1.206(10)(22)[24].
\end{equation}
As a byproduct of our analysis we obtain the values of the pseudoscalar decay constants for the $D_s$ and $D$ mesons as well as for their 
ratio:
\begin{equation}\label{eq:fD_Ds_ratio}
f_{Ds} = 250(5)(5)[7] ~{\rm MeV}, \,\,\,\, f_{D} = 208(4)(6)[7] ~{\rm MeV}, \,\,\,\, f_{Ds}/f_{D} = 1.201(7)(20)[21] 
\end{equation}
Error coding is as in Eq.~(\ref{eq:mb}). 

\section{Computation of bag parameters and $\xi$}
The $\Delta B=2$ effective  weak Hamiltonian in its most general form reads
\begin{equation}
{\cal H}_{\rm{eff}}^{\Delta B=2} =\frac{1}{4}\, \sum_{i=1}^{5} C_i {\cal O}_i 
+\frac{1}{4}\, \sum_{i=1}^{3} \tilde{C}_i \tilde{{\cal O}}_i, \, 
\label{eq:Heff}
\end{equation}
with
\begin{eqnarray}\label{def_Oi}
{\cal O}_1 &=& [\overline{b}^\alpha \gamma_\mu (1-\gamma_5)q^\alpha][\overline{b}^\beta \gamma_\mu (1-\gamma_5)q^\beta],   \,\,\,\,\, 
 {\cal O}_2 = [\overline{b}^\alpha (1-\gamma_5)q^\alpha][\overline{b}^\beta  (1-\gamma_5)q^\beta], \nonumber \\
{\cal O}_3 &=& [\overline{b}^\alpha (1-\gamma_5)q^\beta][\overline{b}^\beta  (1-\gamma_5)q^\alpha], \,\,\,\,\,
{\cal O}_4 = [\overline{b}^\alpha (1-\gamma_5)q^\alpha][\overline{b}^\beta  (1+\gamma_5)q^\beta], \nonumber \\  
{\cal O}_5 &=& [\overline{b}^\alpha (1-\gamma_5)q^\beta][\overline{b}^\beta  (1+\gamma_5)q^\alpha], \,\,\,\,\, 
 \tilde{{\cal O}}_1 = [\overline{b}^\alpha \gamma_\mu (1+\gamma_5)q^\alpha][\overline{b}^\beta \gamma_\mu (1+\gamma_5)q^\beta] \nonumber \\
 \tilde{{\cal O}}_2 &=& [\overline{b}^\alpha (1+\gamma_5)q^\alpha][\overline{b}^\beta  (1+\gamma_5)q^\beta], \,\,\,\,\,\,
\tilde{{\cal O}}_3 = [\overline{b}^\alpha (1+\gamma_5)q^\beta][\overline{b}^\beta  (1+\gamma_5)q^\alpha].   
\end{eqnarray}
In Eqs~(\ref{def_Oi}) $q \equiv d\, {\rm or}\, s$,  $\alpha$ and $\beta$ denote color indices and
spin indices are implicitly contracted within square brackets.
The Wilson coefficients $C_i$ and $\tilde{C}_i$  have an implicit renormalization  scale dependence which
is compensated by the  scale dependence of  the renormalization constants of the corresponding  operators.
The parity-even parts of the operators ${\cal O}_i$ and $\tilde{{\cal O}}_i$ are identical.
Due to parity conservation in strong interactions it is  sufficient
to consider the matrix elements where only the parity-even components of the operators ${\cal O}_i$ enter. 
In the SM only the matrix element of the operator ${\cal O}_1$ enters.
The bag parameters, $B_i$ ($i=1, \ldots, 5$), are defined through the equations 
\begin{eqnarray}
 \langle \overline{B}^{0}_{q} | O_1(\mu) | B^{0}_{q} \rangle &=& {\cal C}_1\, B_1^{(q)}(\mu) ~ 
 m_{B_{q}}^2 f_{B_{q}}^2  \label{B1} \\
 \langle \overline{B}^{0}_{q} | O_i(\mu) | B^{0}_{q} \rangle &=& {\cal C}_i\, B_i^{(q)}(\mu) ~ 
\LSB \dfrac{ m_{B_{q}}^{2} f_{B_{q}}}{ m_b (\mu) + m_{q}(\mu)} \RSB^2  ~~~ {\rm for} ~~ i=2,\ldots, 5, 
\label{Bi} 
\end{eqnarray}
with ${\cal C}_i=(8/3,\, -5/3,\, 1/3,\, 2,\, 2/3)$.
We have used a mixed fermionic action setup to evaluate the four-fermion matrix elements as it  
offers the advantage that matrix elements  are at the same time
O$(a)$-improved and free of wrong chirality mixing effects~\cite{Frezzotti:2004wz}.
Four-fermion RCs have been computed using RI/MOM techniques in~\cite{Bertone:2012cu}.

Since in the static limit each of the five bag parameters is a constant, we can  apply the ratio method adopting a strategy 
analogous to the one used for the $b$-quark mass. 
For more details on the computation of ratios of  bag parameters the reader is referred to 
Ref.~\cite{Carrasco:2013zta}. In Table~\ref{tab:bag_res} we report our results 
in the $\overline{\rm{MS}}$ scheme of Ref.~\cite{mu:4ferm-nlo} at the scale of the $b$-quark mass. 

\begin{figure}[!h]
\begin{center}
\subfigure[]{\includegraphics[scale=0.62,angle=-0]{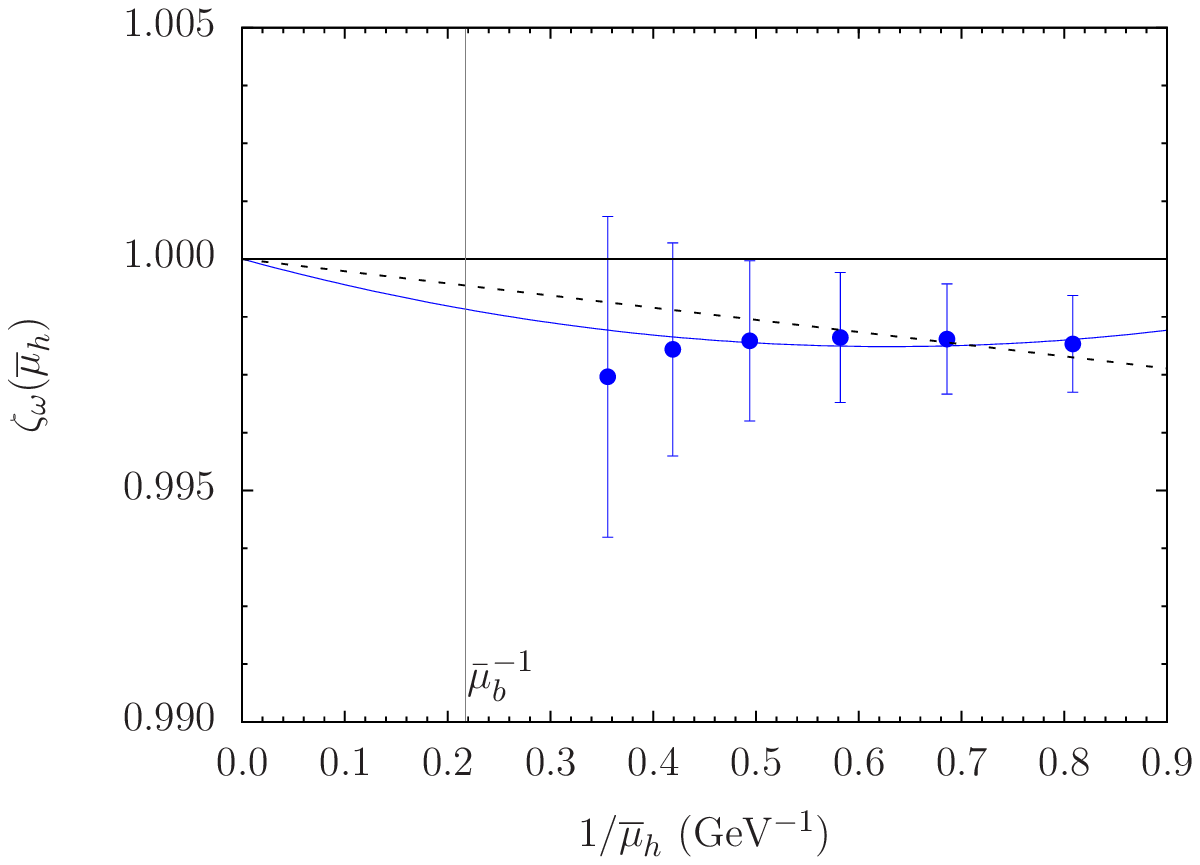}}
\subfigure[]{\includegraphics[scale=0.62,angle=-0]{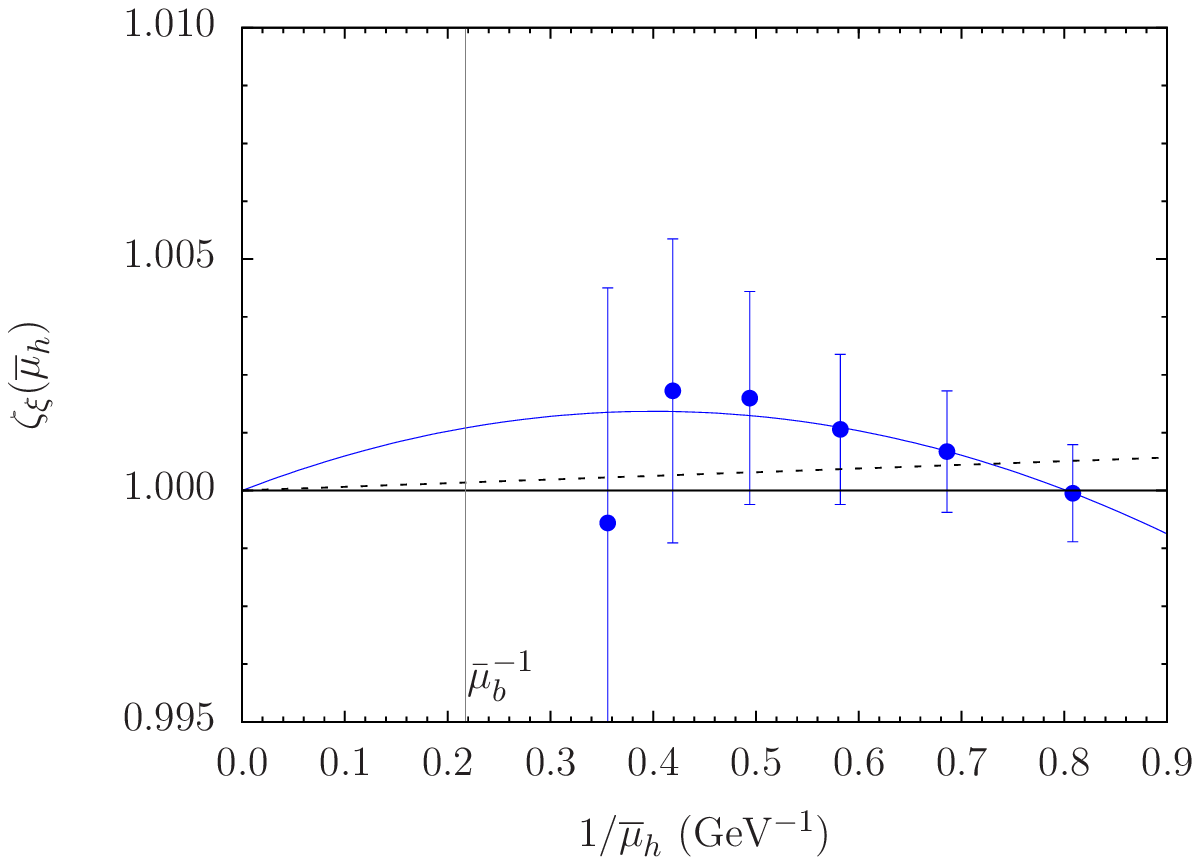}}
\vspace*{-0.3cm}
\caption{ $\zeta_{\omega}(\overline\mu_h)$ and $\zeta_{\xi}(\overline\mu_h)$ against $1/\overline\mu_h$
are shown in panels (a) and (b), respectively.
For both cases the fit function has a polynomial form of the type given by  Eq.~(3.3) (blue curve).
In both panels a fit of the
form $\zeta(\overline\mu_h) = 1 + \eta/\overline\mu_h$ is also performed (black dashed straight line).
The vertical black thin line
marks the position of $1/\overline\mu_b$.
  }
\label{fig:zeta_w_xi_vs_muh}
\end{center}
\end{figure}

\begin{table}[!h]
\begin{center}
\scalebox{0.97}{
\begin{tabular}{|c|c|c|c|c|}
\hline
\multicolumn{5}{|c|}{($\overline {\rm{MS}}$, $m_b$)}\tabularnewline
\hline
\hline
$B_{1}^{(d)}$ & $B_{2}^{(d)}$ & $B_{3}^{(d)}$ & $B_{4}^{(d)}$ & $B_{5}^{(d)}$\tabularnewline
\hline
\hline
0.85(3)(2)[4] & 0.72(3)(1)[3] & 0.88(12)(6)[13] & 0.95(4)(3)[5] & 1.47(8)(9)[12]\tabularnewline
\hline
\hline
$B_{1}^{(s)}$ & $B_{2}^{(s)}$ & $B_{3}^{(s)}$ & $B_{4}^{(s)}$ & $B_{5}^{(s)}$\tabularnewline
\hline
\hline
0.86(3)(1)[3] & 0.73(3)(1)[3] & 0.89(10)(7)[12] & 0.93(4)(1)[4] & 1.57(7)(8)[11]\tabularnewline
\hline
\end{tabular}
}
\caption{Continuum limit results for $B_i^{(d)}$ and $B_i^{(s)}$ ($i=1, \ldots, 5$),
renormalized in the $\overline{\rm{MS}}$ scheme of Ref.~\cite{mu:4ferm-nlo} at the scale of the $b$-quark mass.
Error coding is as in Eq.~(3.5).}
\label{tab:bag_res}
\end{center}
\end{table} 
We also apply the ratio method to compute the SU(3)-breaking ratios $B^{(s)}_1 / B^{(d)}_1$ and the important 
$\xi = (f_{Bs}/f_{Bd})\, (B^{(s)}_1/B^{(d)}_1)^{1/2}$ parameter. 
In Fig.~\ref{fig:zeta_w_xi_vs_muh}(a) and (b) we display 
the dependence on the inverse heavy quark mass of the appropriate ratios for $B^{(s)}_1 / B^{(d)}_1$ and $\xi$, 
denoted as  $\zeta_{\omega}(\overline\mu_h)$ and $\zeta_{\xi}(\overline\mu_h)$, respectively.   
For $B^{(s)}_1 / B^{(d)}_1$ and $\xi$ we obtain  (see Ref.~\cite{Carrasco:2013zta} for details)  
\begin{equation}\label{eq:BB_xi}
 B_1^{(s)}/B_1^{(d)} = 1.007(15)(14)[21], \,\,\,\,\, \xi = 1.225(16)(26)[31].
\end{equation}
Our results for the quantities $f_{Bq}\sqrt{B_i^{(q)}}$ ($i=1, \ldots, 5$ and $q=d/s$) can be found in Table 4 of 
Ref.~\cite{Carrasco:2013zta}.

\section{Model-independent constraints on $\Delta B=2$ operators and NP scale from UTA}

The NP generalisation of the UT analysis is carried out by including matrix elements that, though absent in the SM,  
may appear in the theoretical parametrisation of various observables in extensions of the SM.  
We employ our unquenched lattice QCD results for the bag parameters
of the full basis of the $\Delta B =2$ four-fermion operators to update the UT analysis beyond the SM presented 
in~Ref.~\cite{Bona:2007vi}.
The effective weak Hamiltonian, Eq.~(\ref{eq:Heff}),  is parameterized
by Wilson coefficients of the form $C_i (\Lambda) = (F_i\, L_i) / (\Lambda^2)$ with $i=2,\ldots,5 $. 
We denote by $F_i$ the (generally complex) relevant NP flavor coupling,
by $L_i$  a (loop) factor which depends on the interactions that
generate $C_i(\Lambda)$, and $\Lambda$ is the scale of NP, i.e. the
typical mass of new particles mediating $\Delta B=2$ transitions. 
Hence, the phenomenologically allowed range for
each of the Wilson coefficients can be translated into a lower bound on $\Lambda$.
Other assumptions on the flavor structure of NP correspond to different choices of the $F_i$ functions.

Following Ref.~\cite{Bona:2007vi}, we derive the lower bounds on
the NP scale $\Lambda$ by setting $L_i = 1$ which corresponds to
strongly-interacting and/or tree-level coupled NP. Two other
interesting possibilities are provided by loop-mediated NP contributions with $L_i$
proportional to either $\alpha_s^2$ or $\alpha_W^2$. The first case
corresponds for example to gluino exchange in the minimal
supersymmetric SM, while the latter applies to all models with SM-like
loop-mediated weak interactions. To obtain a lower bound on
$\Lambda$ entailed by loop-mediated contributions, one simply has to
multiply the bounds we quote below by
$\alpha_s(\Lambda)\sim 0.1$ or $\alpha_W \sim 0.03$, respectively.

Our analysis is performed (as in~\cite{Bona:2007vi}) by switching on one
coefficient at a time in each sector, thus excluding the possibility
of having accidental cancellations among  contributions of
different operators.  Thus in the case of accidental cancellations the bounds will be weaker.

In Tables~\ref{tab:Bd} and~\ref{tab:Bs} we collect the 
results for the upper bounds on the $|C^{B_d}_i|$ and $|C^{B_s}_i|$ coefficients and
the corresponding lower bounds on the NP scale $\Lambda$. 
The superscript $B_d$ or $B_s$ is to remind that we
are reporting the bounds coming from the $B_d$- and $B_s$-meson sectors we are here
analyzing. In the  present analysis both experimental and theoretical inputs have been updated with
respect to Ref.~\cite{Bona:2007vi} (see Ref.~\cite{utfitwebpage}). 
The constraints on the Wilson coefficients of 
non-standard operators and, consequently, on the NP scale turn out to be significantly
more stringent than in Ref.~\cite{Bona:2007vi}, in particular for the $B_s$ sector.
Comparing with the results of the UT--analysis in Ref~\cite{Bertone:2012cu}, we notice that
(at least for generic NP models with unconstrained flavour structure) the bounds
on the NP scale coming from $K^0$--$\bar K^0$ matrix elements turn out to be
the most stringent ones.

\begin{table}[!ht]
\parbox{.45\linewidth}{
\centering
\scalebox{0.90}{
\begin{tabular}{|@{}ccc|}
\hline
 & $95\%$ upper limit  &
Lower limit on $\Lambda$ \\
&(GeV$^{-2}$) &
 (TeV)\\
\hline
\phantom{A} $|C^{B_d}_1|$ & $4.7 \cdot 10^{-12}$ & $4.6 \cdot 10^{2}$ \\
\phantom{A} $|C^{B_d}_2|$ & $3.0 \cdot 10^{-13}$ & $1.8 \cdot 10^{3}$  \\
\phantom{A} $|C^{B_d}_3|$ & $1.1 \cdot 10^{-12}$ & $9.5 \cdot 10^{2}$  \\
\phantom{A} $|C^{B_d}_4|$ & $9.5 \cdot 10^{-14}$ & $3.2 \cdot 10^{3}$  \\
\phantom{A} $|C^{B_d}_5|$ & $2.7 \cdot 10^{-13}$ & $1.9 \cdot 10^{3}$ \\
\hline
\end{tabular}
}
\caption {$95\%$ upper bounds for the $|C^{B_d}_i|$ coefficients
  and the corresponding lower bounds on the NP scale, $\Lambda$, for
  a  strongly interacting NP with generic flavor structure ($L_i=F_i=1)$. } 
\label{tab:Bd}
}
\hfill
\parbox{.45\linewidth}{
\centering
\scalebox{0.90}{
\begin{tabular}{|@{}ccc|}
\hline
 & $95\%$ upper limit  &
Lower limit on $\Lambda$ \\
&(GeV$^{-2}$) &
 (TeV)\\
\hline
\phantom{A} $|C^{B_s}_1|$ & $5.6 \cdot 10^{-11}$ & $1.3 \cdot 10^{2}$ \\
\phantom{A} $|C^{B_s}_2|$ & $4.9 \cdot 10^{-12}$ & $4.5 \cdot 10^{2}$  \\
\phantom{A} $|C^{B_s}_3|$ & $1.8 \cdot 10^{-11}$ & $2.3 \cdot 10^{2}$  \\
\phantom{A} $|C^{B_s}_4|$ & $1.6 \cdot 10^{-12}$ & $7.9 \cdot 10^{2}$  \\
\phantom{A} $|C^{B_s}_5|$ & $4.5 \cdot 10^{-12}$ & $4.7 \cdot 10^{2}$ \\
\hline
\end{tabular}
}
\caption {$95\%$ upper bounds for the $|C^{B_s}_i|$ coefficients
  and the corresponding lower bounds on the NP scale, $\Lambda$, for
  a  strongly interacting NP with generic flavor structure ($L_i=F_i=1)$. } 
\label{tab:Bs}
}
\end{table}

\noindent{\bf Acknowledgements}\\
G. H. acknowledges the support by DFG (SFB 1044).
We acknowledge computer time made available to us on the Altix system at the HLRN supercomputing service in Berlin
under the project "B-physics from lattice QCD simulations". 
Part of this work has been completed thanks to allocation of  CPU time on BlueGene/Q -Fermi based on the
agreement between INFN and CINECA and the specific initiative INFN-RM123. 
We acknowledge partial support from ERC Ideas Starting Grant n.~279972 ``NPFlavour'' and ERC
Ideas Advanced Grant n.~267985 ``DaMeSyFla". \\

\end{document}